# Sixty Years of Element Abundance Measurements in Solar Energetic Particles

**Donald V. Reames**
Institute for Physical Science and Technology, University of Maryland, College Park, MD, USA

**Abstract** Sixty years ago the first observation was published showing solar energetic particles (SEPs) with a sampling of chemical elements with atomic numbers $6 \leq Z \leq 18$ above 40 MeV amu$^{-1}$. Thus began study of the direct products of dynamic physics in the solar corona. As we have progressed from 4-min sounding-rocket samples to continuous satellite coverage of SEP events, we have extended the observations to the unusual distribution of element abundances throughout the periodic table. Small "impulsive" SEP events from islands of magnetic reconnection on open magnetic-field lines in solar jets generate huge enhancements in abundances of $^3$He and of the heaviest elements, enhancements increasing as a power of the ion mass-to-charge ratio as $(A/Q)^{3.6}$, on average. Solar flares involve the same physics but there the SEPs are trapped on closed loops, expending their energy as heat and light. The larger, energetic "gradual" SEP events are accelerated at shock waves driven by fast, wide coronal mass ejections (CMEs). However, these shocks can also reaccelerate ions from pools of residual suprathermal impulsive ions, and CMEs from jets can also drive fast shocks, complicating the picture. The underlying element abundances in SEP events represent the solar corona, which differs from corresponding abundances in the photosphere as a function of the first ionization potential (FIP) of the elements, distinguishing low-FIP (<10 eV) ions from high-FIP neutral atoms as they expand through the chromosphere. Differences in FIP patterns of SEPs and the solar wind may distinguish closed- and open-field regions of formation at the base of the corona. Dependence of SEP acceleration upon $A/Q$ allows best-fit estimation of ion $Q$-values and hence of the source plasma temperature of $\sim 1 - 3$ MK, derived from abundances, which correlates with recent measures of temperatures using extreme ultraviolet emission from jets. Thus, element abundances in SEPs have become a powerful tool to study the underlying solar corona and to probe physical processes of broad astrophysical significance, from the "FIP effect" to magnetic reconnection and shock acceleration. New questions arise, however, about the theoretical basis of correlations of energy-spectral indices with power-laws of abundances, about the coexistence of separate resonant and non-resonant mechanisms for enhancements of $^3$He and of heavy elements, about occasional events with unusual suppression of He and about the overall paucity of C in FIP comparisons.

# 1 In Pursuit of Abundances

The Sun is a microcosm for many of the high-energy physical processes that occur throughout the Universe.  Magnetic reconnection generates violent explosive flares and it drives solar jets and liberates coronal mass ejections (CMEs); these CMEs drive collisionless shock waves that ply the solar corona; magnetic turbulence scatters ions differentially, and wave-particle resonances occur throughout.  These processes modify the relative abundances of ions of the available elements in definable ways wherever they occur, including the coronae of distant stars, perhaps even those with habitable planets.  However, in the case of the Sun, unlike distant sources, we have the rare privilege to actually collect direct samples of the energetic ions themselves.  "Multi-messenger" comparisons may help us relate these ions to the limited photons they emit elsewhere.

Measurements of the abundances of the chemical elements has been a rich source of information about the nature of the solar energetic particles (SEPs), their source environment, and of the physical processes that fractionate the solar corona itself, in addition to particle acceleration and transport under a wide variety of circumstances.  We have only to disentangle the individual processes that contribute – a task that has kept us quite busy for 60 years.

The purpose of this article is to acknowledge this milestone, to review the accomplishments, and to highlight remaining questions.

## *1.1 Early Observations, 1961-*

Study of element abundances in SEP events began when Fichtel and Guss (1961) reported the first observation of elements with atomic numbers $Z > 2$.  Their Nike-Cajun sounding rocket flight from Ft. Churchill, Manitoba exposed a nuclear emulsion payload above the atmosphere for several minutes during the SEP event of 3 September, 1960.  The emulsions recorded about two dozen ions of C, N, and O above about 40 MeV amu$^{-1}$ and a few Ne, Mg, Si, and S ions, but an absence of Li, Be, and B.  The author of this review joined that group before the next solar cycle when measurements would be extended up to the abundance peak at Fe, using the same technique (Bertsch et al. 1969).

Comparisons began early of these SEP abundances of elements with those of the solar photosphere, which were also evolving, and Biswas and Fichtel (1964) summarized the abundances of He, C, N, O, Ne, Mg, Si, and S.  With hindsight we might recognize that the heavier elements are somewhat suppressed in these large SEP events, but we would also note their high energy and that the brief rocket shots occurred hours to days after the events began, providing no time histories of the events.

The next decade saw the flight of *dE/dx* vs. *E* telescopes using higher-resolution detectors on satellites.  Early experiments spent most of their time measuring the dominant H and He, but soon, priority systems were invented to insure that ions with $Z > 2$ received a reasonable share of the telemetry space despite high intensities of electrons, protons, and He (e.g. Teegarden et al. 1973).  These measurements showed that abundance ratios, such as C/O, were energy independent from 8 to 40 MeV amu$^{-1}$ and that the abundances of the elements from C through Fe were somehow comparable with photospheric or coronal abundances, as far as these were known. They also disagreed strongly with a previous event measured by Mogro-Campero and Simpson (1972) that showed a smooth systematic increase in the abundance enhancement relative to the corona by a factor of order ten vs. *Z* between C and Fe. These authors recognized the enhancement as a probable dependence on the mass-to-charge ratio *A/Q* of the ions. Did SEP abundances have a strong dependence on *A/Q* or not?

Figure 1 compares samples of the evolution in particle resolution and measurement. Fig 1a shows the data of Fichtel and Guss (1961), Fig. 1b shows resolution up to Fe by Bertsch et al. (1969), and Fig. 1c shows resolution typical of Si, solid-state detectors flown on satellites beginning in the 1970's. It is the accumulation of ions over 43 large SEP events for the average abundance study by Reames (1995a).

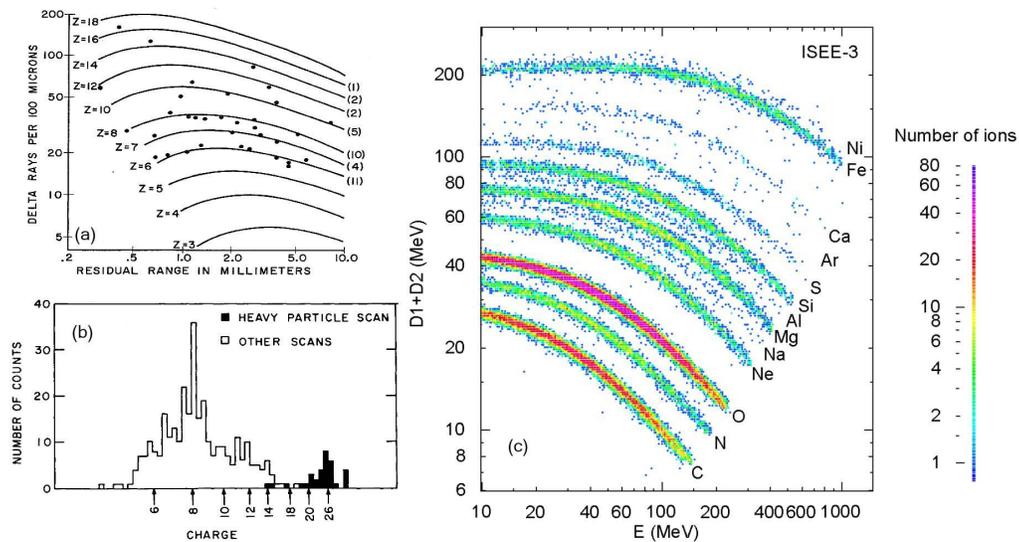

**Fig. 1** Panel **(a)** shows identification of ions by counts of δ rays (scattered electrons) per unit length along particle tracks in nuclear emulsions flown on sounding rockets by Fichtel and Guss (1961), **(b)** shows similar resolution of elements up to Fe by Bertsch et al. (1969), and **(c)** shows resolution of Si, solid-state detectors with data accumulation during 43 large SEP events from 1978 – 1988 in a histogram of *dE/dx* vs. *E* by Reames (1995a). SEP abundances are now measured by simply counting ions of each element in a constant velocity interval.

## *1.2 $^3$He-rich Events, 1970-*

Nearly every scientist who considered SEPs had previous experience with galactic cosmic rays (GCRs) and the lessons learned from GCRs would all be tested against SEPs (often called SCRs for solar cosmic rays). GCRs are accelerated by shock waves from supernovae and during their ~10$^7$-year sojourn in interstellar space they undergo nuclear collisions with interstellar H, producing significant abundances of secondary ions such as $^2$H and $^3$He and isotopes of Li, Be, and B. Early abundances of SEPs had shown a lack of Li, Be, and B. However, when Hseih and Simpson (1970) observed an SEP event with $^3$He/$^4$He = 0.021±0.004, much enhanced compared with 4×10$^{-4}$ in the solar wind (e.g. Gloeckler and Geiss 1998), it seemed to be possible evidence of nuclear reactions in solar flares after all, especially since γ-ray lines had also been observed from flares (Chupp et al. 1973; Ramaty and Murphy 1987). However, subsequent observations were soon found of many events like that seen by Serlemitsos and Balasubrahmanyan (1975) where $^3$He/$^4$He = 1.52 ± 0.10 but $^3$He/$^2$H > 300. These huge enhancements of $^3$He with no $^2$H or $^3$H were incompatible with nuclear reaction products. The case closed further when limits on Be/O or B/O in large SEP events were found to be < 2 × 10$^{-4}$ (e.g. McGuire et al. 1979; Cook et al. 1984). Reaction products are indeed produced in flares, but flares occur on closed magnetic loops so ions cannot get out. SEPs in space do not come from flares, and we need a whole new resonance mechanism for these "impulsive" SEP events.

There followed many years of many proposals for the enhancement of $^3$He, first from preferential preheating by resonant wave absorption (e.g. Ibragimov and Kocharov 1977; Kocharov and Kocharov 1978, 1984; Fisk 1978; see Sect. 2.5.2 of Reames 2021b and references therein), which required additional acceleration. Generally, waves that resonate with the gyrofrequencies of the much-more-abundant H and $^4$He are absorbed by them during stochastic acceleration, but $^3$He with a gyrofrequency between these two dominant ions continues to absorb waves that it is too rare to damp. Essentially, while H and $^4$He assume nearly power-law spectra, nearly all of the $^3$He in the active volume gets accelerated to a peak in the region ~0.1 – 1 MeV amu$^{-1}$ (e.g. Liu et al. 2006). Temerin and Roth (1992) suggested preferential acceleration by absorption of electromagnetic ion cyclotron (EMIC) waves generated by streaming electrons. Impulsive SEP events had been shown to be accompanied by the intense beams of electrons (Reames et al. 1985) that produce type III radio bursts (Reames and Stone 1986). Temerin and Roth (1992) envisioned $^3$He ions absorbing incoming waves as they mirrored in magnetic fields, in analogy with ion conics seen in the Earth's aurorae. Their model led to acceleration, not just preheating, and associated the $^3$He with the electrons.

It also became clear that the element abundances in $^3$He-rich events showed a systematic enhancement with Z that grew to a factor of ten for Fe/O (e.g. Mason et al. 1986; Reames et al. 1994; Mason 2007) when abundances in these small events were compared with average abundances in large "gradual" SEP events. A study of daily-averaged abundances over an 8.5-year period showed the existence of two distinct populations of particles (Reames 1988). On a plot of daily intensities of Fe vs. O at ≈2 MeV amu$^{-1}$, the Fe-rich branch was also $^3$He-rich, electron-rich, and proton-poor. A modern version of this plot in Fig. 2 shows intensities of Fe vs. O with $^3$He/$^4$He enhanced in color and point size in the upper panel and heavy element abundances (50 ≤ Z ≤ 56)/O enhanced in the lower panel.

Observations of γ-ray lines in large events also began to show evidence of enhancements in the Doppler-broadened lines of the accelerated Fe/O (Murphy et al. 1991) and $^3$He/$^4$He (Mandzhavidze et al. 1999; Murphy et al. 2016), suggesting the same physics in flares as in the impulsive SEP events we see in space. Flares were understood to involve magnetic reconnection on closed field lines, but impulsive SEP events were not initially associated with jets, which involve reconnection on open field lines.

Thus, $^3$He-rich events were a major discovery that focused the mind on sources and physical mechanisms. Earlier theory of SEPs had been focused almost exclusively on diffusive transport with very little focus on acceleration (e.g. see Chap. 2 of Reames 2021b).  If $^3$He-rich events were somehow related to solar flares, what was the origin of the "big proton events"?  Why were they *not* $^3$He rich?

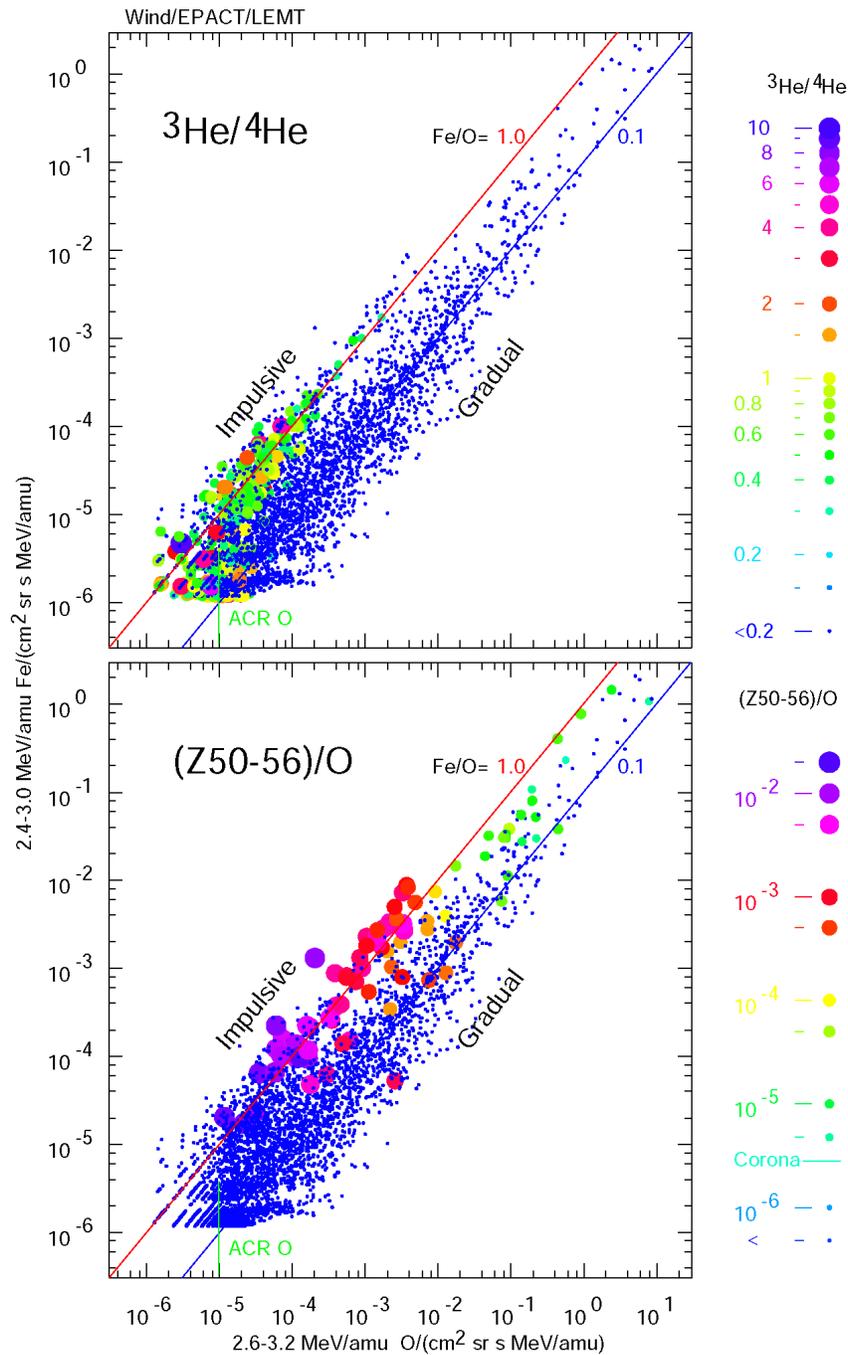

**Fig. 2**  In each panel, particle intensities of Fe are plotted vs. intensities of O for all 8-hour intervals in an 8.5-year study period from 1994 through 2003 whenever Fe and O were measurable. The indicated intensity of anomalous-cosmic-ray (ACR) O forms a lower bound on the O intensity during solar minimum. The symbols used for points in the upper panel indicate the average value of $^3$He/$^4$He during each interval, as indicated in the scale to the right of the panel. Symbols in the lower panel denote the (50≤Z≤56)/O abundance ratio shown in the right-hand scale. Lines drawn along Fe/O =1 and 0.1, approximately track the locus of impulsive and gradual SEP events, respectively (Reames and Ng 2004).  This is an updated version of the bimodal abundance study of Reames (1988).

Following a brief earlier comparison by Reames (1999), it was found that as event sizes increased, $^3$He became depleted from its probable source volume and had a limited fluence (Ho et al. 2006).

## 1.3   Ionization States, 1984-

Luhn et al. (1984, 1987) provided the earliest direct measurements of ionization states, $Q$, of the ions of elements up to Fe for energies 0.34 – 1.8 MeV amu$^{-1}$. For gradual events they found average values of $Q_{Si}$ = 11.0 ± 0.3 and $Q_{Fe}$ = 14.1 ± 0.2 suggesting a temperature near ≈ 2 MK. However, for $^3$He-rich events they found $Q_{Si}$ ≈ 14 and $Q_{Fe}$ = 20.5 ± 1.2, which could suggest either a source temperature of ≈10 MK, or stripping of the ions after acceleration. Generally, the results were initially misinterpreted as the ions in impulsive events coming from hot flares. It was not initially recognized how difficult it would be to enhance abundances of any of the elements Ne, Mg, and Si relative to He, C, N, and O, if all of them were fully ionized with $A/Q$ = 2 at 10 MK, suggesting that they were actually accelerated at some lower temperature, then stripped by traversing a small amount of matter afterward.

Additional measurements of $Q$ in gradual events, including those that used relative deflection of ions in the Earth's magnetic field, measured ions with energies up to 200 - 600 MeV amu$^{-1}$ and showed that $Q_{Fe}$ varied from 11 to 15 (Mason et al 1995, Leske et al. 1995, 2001; Tylka et al 1995; Klecker 2013) in different events.

The alternate interpretation of the ionization-state measurements in impulsive events became a requirement when DiFabio et al. (2008) found that the charges increased with particle energy, suggesting that the ions were increasingly stripped to a higher $Q$ at higher speed, in fact, attaining an equilibrium $Q$ appropriate for each speed. Thus these ions were clearly stripped after acceleration when they traversed a small amount of matter. This might occur for acceleration at a depth of about 1.5 $R_S$ in the corona, the authors suggested.

A different type of logic was applied by Reames et al. (1994) to determine ionization states. They noted that in impulsive events, on average, the elements He, C, N, and O had the relative abundances expected from a sample of the corona, Ne, Mg, and Si, were each enhanced by a factor of about 2.5, and Fe was enhanced by a factor of ≈7. They thought this suggested that He through O were all fully ionized with $Q/A$ = 0.5, suggesting a temperature of 3 – 5 MK; above ~3 MK was required to keep O ionized. The similarity of the enhancements for Ne, Mg, and Si suggested that they were each in a highly stable state with two orbital electrons so that $Q/A$ ≈ 0.43, as only occurs for $T$ < 5 MK, above which Ne loses electrons. This would later provide the basis for using abundances enhancements to estimate temperatures. At the time it was also another argument against the possibility of ~10 MK solar flares as a source.

## 1.4   FIP and A/Q,  1975-

It was recognized early that SEP abundances, especially in large (gradual) events, had a dependence on the first ionization potential (FIP) of the elements, relative to the corresponding photospheric abundances. In the photosphere, elements with low FIP (<10 eV) were ionized while those with high FIP were neutral atoms and unaffected by electromagnetic fields. A similar FIP dependence had been suggested for the GCRs, compared with "universal" abundances, and this strong correspondence was noted early by Webber (1975).

Meyer (1985) reviewed the increasingly extensive SEP abundance measurements and characterized the events (excluding $^3$He-rich events) as having a common "un-biased baseline" that differed from the photosphere as a function of FIP, and a "mass bias" dependence that was actually a function of $A/Q$ and varied from event to event. Presumably the baseline SEP abundances represented the abundances in the corona, determined long before acceleration, while the $A/Q$ dependence might occur during acceleration. We do not need to decide between FIP- and $A/Q$-dependence of SEPs, we need both.

Meanwhile, Breneman and Stone (1985) used the new average $Q$-values from Luhn et al. (1984) to plot abundance enhancements vs. $A/Q$ and actually showed a power-law dependence that could increase with $A/Q$ in some events and decrease in others. Of course, one would not expect all SEP events to come from a single temperature environment and have the same pattern of $Q$ values as the Luhn et al. (1984) average, but $A/Q$ probably represents the residual affects of ion magnetic rigidity, and one could easily expect physical processes of SEP acceleration and transport that vary as power laws in magnetic rigidity. Thus, if we want to understand the abundance pattern in an SEP event, we must know the power-law $A/Q$ pattern of the elements, i.e. we must know an effective temperature for the source plasma contributing the ions with appropriate $Q$ values in that event. For SEPs from the corona, a temperature determines the $Q$-values and the power law in $A/Q$ determines the abundances enhancements (e.g. Reames 2018b).

Breneman and Stone (1985) also believed that correction by a power-law in $A/Q$ was needed between average SEP and coronal abundances. More-recent SEP abundances averaged over many large events are usually treated directly as coronal abundances (Reames 1995a, 2014). This is based partly on the idea that differences in ion transport will cause different spatial distribution of different elements, but these will average out when we integrate over many events observed from diverse perspectives. The elements Mg and Si have nearly the same FIP as Fe, but different values of $A/Q$; yet, they have similar relative enhancements on a plot of FIP dependence.

Isotope resolution, for elements other than He, have been extended up to Fe (e.g. review by Leske et al. 1999, 2007). Basically, these measurements show variations in $A/Q$ for single elements that are similar in chatacter to those seen more broadly for different elements (e.g. Breneman and Stone, 1985).

## 1.5   *Self-generated Waves*

Early work ascribed nearly all aspects of space-time variation in SEP events to scattering during transport from a point source with small scattering mean free paths ($\lambda \sim 0.1$ AU), the "Palmer (1982) consensus," which were treated as invariant, pre-existing properties of interplanetary space. Then Mason et al. (1989) found that $^3$He-rich events were nearly scatter free ($\lambda \geq 1$ AU), even when they occur in the slowly declining phase of a large gradual SEP event (e.g. Chap. 2 in Reames 2021b).  If the slow decline in intensity is caused by intense scattering, how can the newly-injected ions travel scatter free?

At the other extreme, the largest gradual events showed evidence of proton amplified waves that differentially affected ions, altering abundance ratios, limited intensities early in events, and flattened low-energy spectra (Reames 1990, Reames et al. 2000, Ng and Reames 1994, Ng et al. 1999, 2003, 2012; Reames and Ng 2010).  When intense, streaming protons are present, scattering can become a strong function of rigidity, space, and time.

## 2  The Turn of the Century

As we approached the year 2000, we had fairly extensive observations of SEPs at locations throughout the heliosphere.  Two major classes of SEP events had been resolved, called impulsive and gradual, where ion abundances suggested different physics dominated particle acceleration.  The origin of gradual SEP events seemed most clear; particles in gradual events were accelerated at shock waves (e.g. Lee 1978, 2005) driven by fast, wide CMEs (Kahler et al. 1984) and they could rapidly reach GeV energies (Zank et al. 2000, 2007; Cliver et al. 2004; Sandroos and Vainio 2007; Ng and Reames 2008).  Kahler et al. (1984) found a 96% correlation between large gradual SEP events and fast, wide CMEs; Mason et al. (1984) found consistent abundance ratios over broad longitude intervals that would not support ideas of cross-field transport (see many reviews e.g by Reames 1999, 2013, 2021b; Lee et al. 2012; Desai and Giacalone 2016). Impulsive events were loosely associated with magnetic reconnection and solar flares involving stochastic acceleration (Ramaty 1979; Miller et al. 1997; Miller 1998). The extreme enhancement of $^3$He required resonant wave-particle interactions, as discussed above, possibly related to the associated streaming electrons that produced type III radio bursts (Reames 1985; Reames and Stone 1986).  Actually, the first suggestion of two classes of SEP events had been based upon radio bursts (Wild et al. 1963), type III bursts were produced by "pure" streaming electrons while type II bursts took place at shock waves where proton acceleration was known to occur.  These "pure" electron events (Lin 1970, 1974) turned out to actually be $^3$He-rich events (Reames et al. 1985).

The final decade of the 1990s had surfaced an extensive controversy between the SEP community and physicists who still assumed that all SEPs must come from flares, especially energetic SEPs that could threaten astronauts.  Gosling's (1993, 1994) paper entitled "The solar flare myth" reiterated the importance of shock waves driven by CMEs, not of flares, for the largest gradual SEP events, and reviewed evidence for two classes of SEP events.  This paper was seen to "wage an assault on the last 30 years of solar-flare research" (Zirin 1994).  Three perspectives on the controversy were invited by *Eos*: Hudson (1995) argued that the term "flare" should include the CME, shock, and any related physics; Miller (1995) argued that flares were more numerous and thus better subjects for acceleration studies; Reames (1995c) argued that distinguishing the physics of all sources, especially their spatial extent, was important for SEPs.  The next decades, described below, would improve our estimates of SEP source properties, and would implicate solar jets, with magnetic reconnection involving open field lines, as the actual source of the impulsive SEP events seen in space (Kahler et al. 2001; Bučík 2020).

From a modern perspective, it turns out that as much as half of the energy of magnetic reconnection can directly produce energetic particles (Aschwansen et al. 2019), especially electrons. When trapped on closed loops, these SEPs soon scatter into the magnetic loss cone to deposit their energy in the denser corona below, where the heating causes hot, bright plasma to expand back up into the loops – a solar flare. Thus it may be more accurate to say that SEPs cause flares than the converse. Flares exist precisely because these SEPs and the heated, >10 MK plasma are well trapped.

Of course, this evolution of our understanding of SEPs was not only driven by abundance measurements. Spatial distributions of protons (e.g. Reames et al. 1996, 1997), correlations with CMEs (Kahler 2001; Koulumvakos et al. 2019; Rouillard et al. 2011, 2012, 2016), onset timing (Kahler 1994; Tylka et al. 2003; Reames 2009a, b), electron/proton ratios (Cliver and Ling 2007; Cliver 2016) and intense events and spectral breaks (Tylka and Dietrich 2009; Mewaldt et al. 2012; Gopalswamy et al. 2012) have contributed and are discussed in many reviews (Reames 1999, 1995b, 2013, 2015; Mason 2007; Lee et al. 2012; Desai and Giacalone 2016) and even a textbook on SEPs (Reames 2021b). Nevertheless, observations of element abundances continued to grow and contribute significantly.

## 2.1    Impulsive SEP Events

### 2.1.1 Heavy Elements

Abundances of elements significantly above Fe in the periodic table had been first reported in an early measurement by Shirk and Price (1974) who studied etch pits in the glass window of the *Apollo 16* command module from the small SEP event of 18 April 1972. They found the enhancement of $(Z > 44)/Fe = 120^{+120}_{-60}$, relative to solar abundances, at $0.6 \leq E \leq 2.0$ MeV amu$^{-1}$. Years later, using data from the *Wind* spacecraft for a five-year period Reames (2000) found many impulsive events with enhancements in $(50 \leq Z \leq 56)/Fe$ ranging from 50 to 1000 for energies of $3.3 \leq E \leq 10$ MeV amu$^{-1}$, with no significant enhancement in gradual SEP events. Similar results were also found for $(70 \leq Z \leq 82)/Fe$ with poorer statistics. A comparison of the abundance maxima at Fe, $34 \leq Z \leq 40$, and $50 \leq Z \leq 56$, relative to O, showed enhancements of ~10, ~100, and ~1000 times coronal values, respectively, for impulsive SEP events. These groups of $Z$ we consider correspond to relative maxima of element abundances vs. $Z$ across the periodic table.

With increasing data, Reames and Ng (2004) found the pattern of correlation of $(50 \leq Z \leq 56)/O$ values with Fe/O shown in Fig. 2. To continue abundance measurements up to Pb, it was actually easier to measure the average enhanced value in small impulsive events than the un-enhanced reference value in much larger gradual events, so the coronal reference was estimated from a FIP-corrected photospheric value in Reames and Ng (2004), but an average could be determined for gradual events with additional data ten years later. Thus, Reames et al. (2014a) found a power-law correlation of enhancement vs. $A/Q$ at ≈3 MK with a power of $3.64 \pm 0.15$ shown in Fig. 3b. An enhancement of ≈900 was seen for $(76 \leq Z \leq 82)/O$ relative to measurements in gradual SEP events as a coronal proxy. Below 1 MeV amu$^{-1}$, Mason et al. (2004) measured a factor of ≈200 enhancement for ions with masses $180 \leq A \leq 220$ and a power-law slope vs. $A/Q$ of 3.26 for the average abundances shown in Fig. 3a. The heavy-element enhancements in impulsive SEPs are nearly as extreme as those in $^3$He, but they are uncorrelated with those in $^3$He, suggesting that they involve a different mechanism. These instruments lack resolution of individual elements at $Z > 30$ but are quite adequate to show the extreme enhancement using element groups.

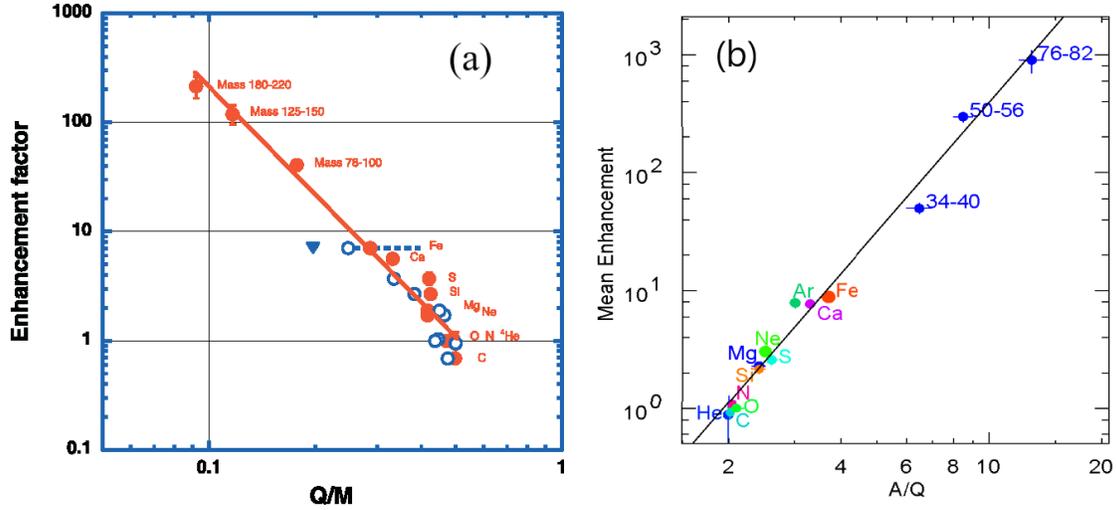

**Fig. 3** Panel **(a)** shows abundance enhancement vs. $Q/M$ (our $Q/A$) at 3.2 MK for named elements and mass intervals at $0.1 \leq E \leq 1.0$ MeV amu$^{-1}$ (*red*: average of events; *blue*: three large events; Mason et al. 2004); Panel **(b)** shows abundance enhancement vs. $A/Q$ at ≈3 MK for named elements and intervals of $Z$ at $3.3 \leq E \leq 10$ MeV amu$^{-1}$ (Reames et al. 2014; Reames and Ng 2004). Departures reflect some significant variations in individual events, especially at low energies.

### 2.1.2 Temperatures

The demonstration by Breneman and Stone (1985) of power-law dependence of abundance enhancements on $A/Q$ lay dormant for many years. To study power-laws in $A/Q$ we must have reasonable estimates of $Q$ or a temperature that allows us to determine $Q$ values. Following the suggestion of Reames et al. (1994) that abundances imply temperatures between 3 and 5 MK, and, later, that average abundance enhancements from He to Pb fit a power law in $A/Q$ with $Q$ values determined near 3 MK (Reames et al 2014a), it seemed reasonable to try to fit the abundance enhancements for each impulsive SEP event using each temperature on a grid of temperatures spaced by 0.1 in log $T$, from 2 to 5 MK (Reames et al. 2014b). The best fit for each event was chosen as the $T$ value that gave the minimum value of $\chi^2$ for the fit. For 79 of the 111 impulsive events, the best fit fell at $T = 2.5$ MK, with another 29 events at 3.2 MK. This study had initially selected impulsive events as those having Fe/O values over four times the reference coronal value; loosening this criterion and including up to 30% systematic abundance errors, in addition to statistical errors (Reames et al. 2015; see also review Reames 2018b), very few new impulsive events outside the 2 – 4 MK region were found. A unique feature in impulsive SEP events is that abundance enhancements tend to follow the order Ne > Mg > Si, opposite the order of $Z$; this is also a unique feature that matches the order of $A/Q$ in the region around $T \approx 2.5$ MK (Reames 2018b). If the ions are stripped after acceleration (DiFabio et al. 2008), abundances provide the only way to determine source plasma temperatures of the ions before or during acceleration.

Good agreement with these source temperatures has recently been found for values of 2 – 2.5 MK, obtained using the differential-emission-measure technique for determining temperatures from the extreme ultraviolet (EUV) images of coronal source regions of 24 solar jets associated with $^3$He-rich events (Bučík et al. 2021). Solar magnetic fields can efficiently insulate regions of differing temperature and density on a fine spatial scale; reconnection in jets is complex, with open-field structures at ~2 MK that emit in the EUV, while nearby closed loops at ~10 MK emit X-rays simultaneously.

### 2.1.3 Acceleration in Jets

For many years, CMEs were not seen with small impulsive SEP events. Then Kahler et al. (2001) observed small, narrow CMEs associated with several of the larger impulsive SEP events. Thus the lack of association could have been only an issue of the CME detection threshold. The narrow CMEs and the previous strong association of impulsive SEPs with type III radio bursts suggested a source of acceleration in solar jets where magnetic reconnection was occurring on open field lines (Kahler et al. 2001). Jets had been considered as a source of type III bursts by Shimojo and Shibata (2000). SEP events could be traced back to their sources (Nitta et al. 2006; Wang et al. 2006; Ko et al. 2013; Reames et al. 2014a) and eventually impulsive events were associated routinely with individual solar jets (Bučík et al. 2018a, 2018b, 2021; see review Bučík 2020).

The theory of magnetic reconnection was first seen as an explanation for the power-law enhancements of heavy elements by Drake et al (2009). These particle-in-cell simulations of the shrinking islands of magnetic reconnection found that ions were Fermi-accelerated as they reflected back and forth from the ends of collapsing islands. This acceleration is able to overwhelm any beta-deceleration from the decreasing fields, and it depends upon the strength of the guide field, out of the plane of the reconnection. These considerations also apply to electron acceleration in reconnection regions (Arnold et al. 2021). These studies do not discuss a role for $^3$He, but the islands of reconnection certainly provide many sites for mirroring ions envisioned by Temerin and Roth (1992).

A sketch of the jet created by emerging flux is shown in Fig. 4. The reconnection does not take place in a single point, but in a series of islands, and the opposing fields rarely cancel exactly, leaving a remnant "guide field." The particles, mainly electrons, are Fermi-accelerated as they pitch-angle scatter in the changing fields (e.g. Arnold et al 2021). As the SEPs and CME are escape along open fields to the upper right in Fig.4, newly-closing field loops are forming on the lower left in the region labeled "Flare"; as these loops close they capture SEPs which deposit their energy as heat (10 MK) or the electrons emit X-ray Bremstrahlung as they scatter against ions in the denser plasma. While most of the reconnection region involves 2 – 3 MK plasma that emits in the EUV (Bučík et al., 2021), even 2 – 3 MK SEPs (Reames et al., 2014b), jets must also have a region of newly-forming flare loops at >10 MK that emit X-rays. As some field lines open, others must close. It is interesting to note, that as the energy in the reconnection grows, both the SEPs and the flaring will grow, but their correlation does *not* mean that flares cause SEPs, rather this is what Kahler (1982) called "big-flare syndrome." Someday we will have overlaid maps of the X-ray and EUV regions that help us understand the detailed structure of a jet.

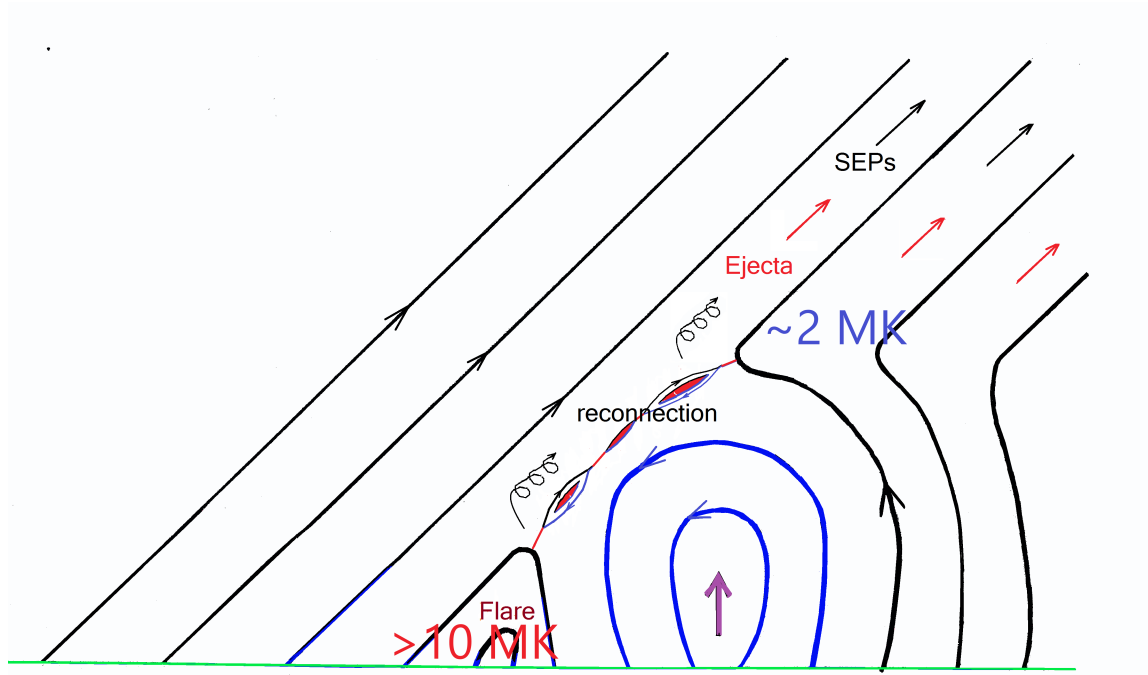

**Figure 4.** A jet is produced when newly emerging magnetic flux (*blue*) reconnects with oppositely directed field (*black*) in the *red* reconnection region. This region is not a uniform surface but forms turbulent islands of reconnection; the fields do not perfectly cancel. SEPs and plasma can escape along *B* toward the *upper right* and a newly-enclosed flaring region labeled "Flare" forms at the *lower left* which emits X-rays. Real jets can be much more complex, involving twisted fields, etc. (Reames 2021b)

If CMEs accompany impulsive SEP events, some of them will be fast enough to drive shock waves that can accelerate or reaccelerate the SEPs. Those shock waves are capable of accelerating ions from the ambient plasma as well as those $^3$He-rich, Fe-rich suprathermal ions from the reconnection, i.e. they have a two-component seed population. When Reames (2019b) sought to extend the *A/Q* power laws of impulsive SEP events down to H at *A/Q* = 1, he found that in some events, usually small, H seemed to fit perfectly on the same power law as ions with *Z* >2, but in many, larger impulsive events, H was a factor of ~10 above the extended power law. This "proton excess" was strongly associated with event size and with the presence of fast CMEs. Events with no proton excess were classified SEP1 events (Reames 2020a) while those with a large proton excess were called SEP2 events (Reames 2020a). Examples of each are compared in Fig. 5.

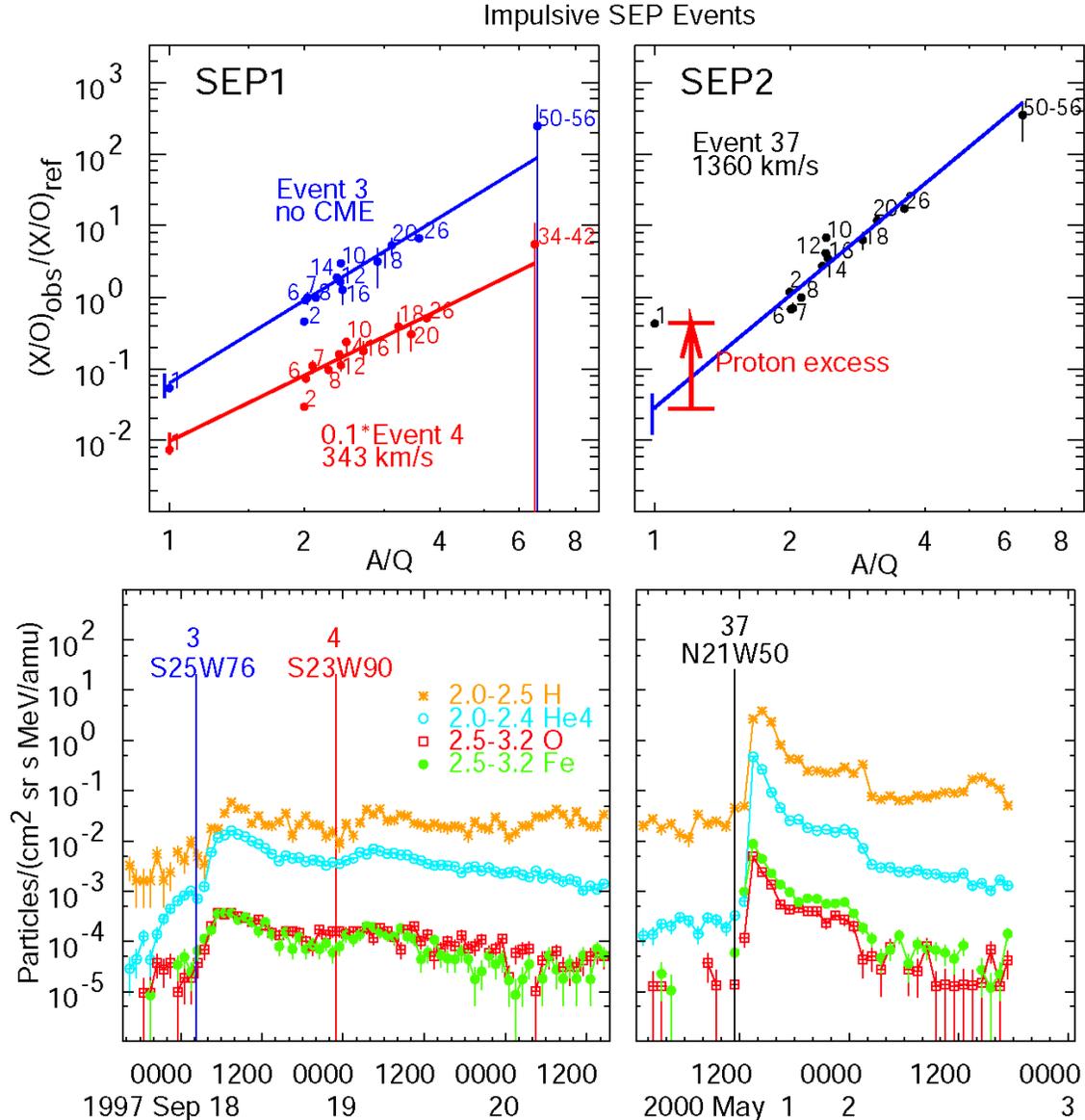

**Fig. 5** *Lower* panels show time histories of H, ⁴He, O and Fe, at the indicated MeV amu⁻¹, for three impulsive SEP events, two small SEP1 events on the *left* and a larger SEP2 event on the *right*. Event numbers shown above source coordinates refer to the event list of Reames et al. (2014a). Power-law fits to the abundance enhancements, noted by $Z$, in each event, are shown in the *upper* panels and CME speeds are listed when CMEs are seen. The *right*-hand Event 37 is an SEP2 class event because of its large proton excess noted. For the SEP1 events on the *left*, the protons abundances lie on the extrapolated power-law fits (Reames 2019b, 2020a).

The lower panel in Fig. 6 shows how larger impulsive events have large proton excesses and fast CMEs while small impulsive SEP events tend to lack both. The upper panel shows how ions with $Z \geq 2$ might be reaccelerated by a shock wave out of pre-enhanced impulsive suprathermal (SEP1) ions in its seed population to dominate high $A/Q$, while the H is dominated by seeds from the ambient coronal ions which are mildly suppressed at high $A/Q$. For impulsive events the proton excess seems to be a signature of reacceleration by shock waves; SEP2 events involve shocks while SEP1 events do not.

**Fig. 6** Panel **(a)** shows the peak proton intensity at 2 – 2.5 MeV vs. the proton excess relative to the $Z >2$ power-law fit for 90 impulsive SEP events with the symbol size and color determined by CME speed as shown. Panel **(b)** suggests two possible contributions, impulsive SEP1 seed particles (*blue*) and ambient coronal seed particles (*red*), to the plot of enhancement vs. $A/Q$ for the shock-enhanced SEP2 impulsive events.

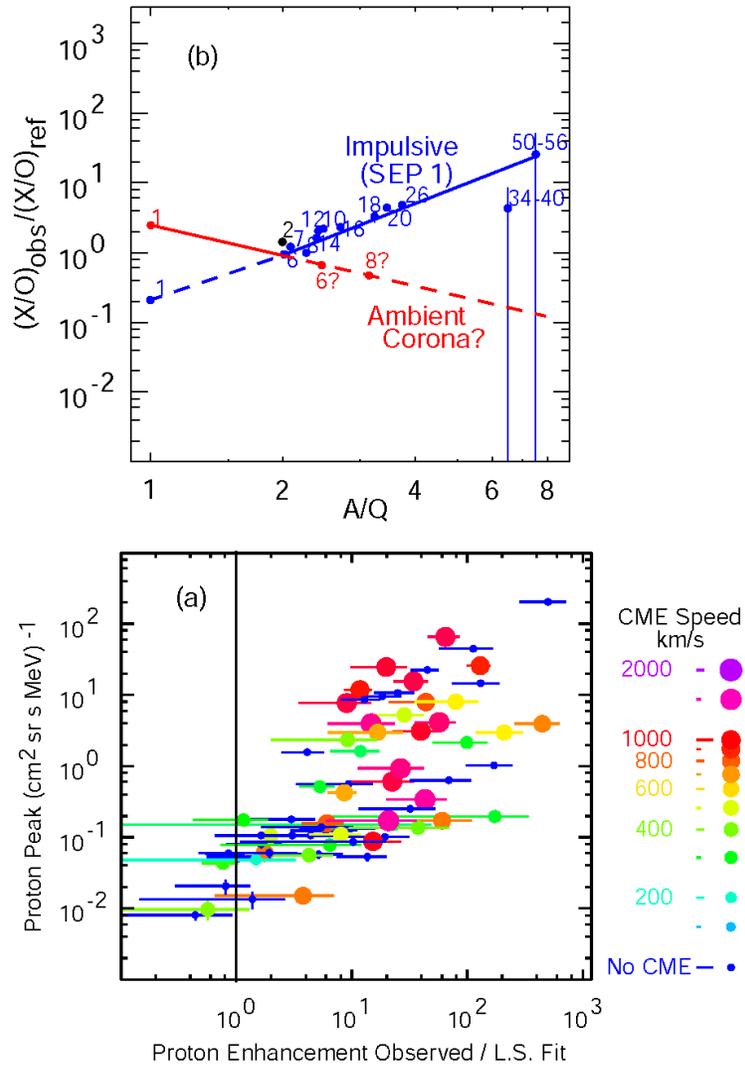

## 2.2  Gradual SEP Events

### 2.2.1 Seed Populations

There was a brief period when the SEP sources and abundances seemed simple: the small impulsive SEP events were $^3$He-rich and the large gradual SEP events were not. Then Mason et al. (1999) found a small increase in $^3$He in a large SEP event that was otherwise gradual in every respect. The increase was only five times the solar-wind average, but it was definite. It was immediately clear that shock waves in gradual events could easily re-accelerate any remnant suprathermal ions left over from previous impulsive events; in fact some shocks might even favor these pre-accelerated ions that are initially fast enough to catch up with the shock from behind even when the magnetic field lies near the plane of the shock in the "quasi-perpendicular" case (Tylka et al. 2001, 2005; Tylka and Lee 2006). Pre-accelerated impulsive ions solve the "injection problem" (Zank et al. 2001) in this case.

   For many years we have observed that the SEP element abundances during quiet periods between events during solar maximum have persistently high abundances of $^3$He/$^4$He and Fe/O (Richardson et al. 1990). These ions are believed to come from many small impulsive SEP events that are too small to be resolved individually. Subsequent observations with more-sensitive instruments have extended these observations (Desai et al., 2003; Wiedenbeck et al. 2013; Bučík et al., 2014, 2015; Chen et al., 2015) suggesting pools of $^3$He-rich, Fe-rich ions that are available in some locations for acceleration by passing shock waves (Reames 2016b).

## 2.2.2 Temperatures

If we intend to study dependence of abundances on *A/Q*, we must have a reasonable estimate for values of *Q* that may change from event to event, possibly even with time during an event. As a first approximation, we can assume that the source is isothermal; that gives us a theoretical estimate of *Q* for each element (e.g. Mazzotta et al. 1998) and allows for event-to-event variation and possible variation of the ion source with time during an event (such time variations have not been seen). Thus we can apply the same technique to gradual events that we used in impulsive events, i.e. at each time interval, we consider all temperatures of likely interest, plot the abundance enhancement vs. *A/Q*, using the *Q* values for that *T*, and determine the fit and its $\chi^2$ value; then we choose the *T* and fit with the minimum $\chi^2$ (Reames 2016a, 2018b).

   Best fits for most (69%) of the gradual events fall in the range $0.8 \leq T \leq 1.6$ MK, while 24% had $2 \leq T \leq 4$ MK, like impulsive events. This difference was attributed to differences in the seed populations: most gradual events, especially large events, are dominated by shock acceleration of ambient coronal ions, but those with $2 \leq T \leq 4$ MK are dominated at $Z > 2$ by pre-accelerated impulsive suprathermal ions in the seed population. The latter events tend to involve weaker or quasi-perpendicular shock waves that benefit from the chance presence of pools of $^3$He-rich, Fe-rich pre-accelerated seed ions described in the previous section. Reames (2020a) distinguishes the events dominated by reaccelerated impulsive SEPs as SEP3 events and those dominated by ambient coronal plasma as SEP4 events. When H is compared with the *A/Q* power laws, the SEP3 events tend to have large proton excesses, while H usually fits the power law in the SEP4 events, as shown for the two SEP4 events in Fig. 7.

   For each 8-hr period shown in Fig. 7c a fit to abundance enhancements relative to SEP-reference abundances (e.g. Table 2) vs. *A/Q* is determined at each temperature, *T*, and the $\chi^2/m$ of the fit is plotted vs. *T* in Fig. 7d. The minimum $\chi^2$ selects the value of *T* shown in Fig. 7c and the best fit power-law vs. *A/Q* shown in Fig 7e. For the two large SEP4 events analyzed in Fig. 7, proton abundances lie near the best-fit power laws for the ions with $Z \geq 6$, suggesting that all elements in each event are derived from the same seed population, i.e. the ambient corona at $T \approx 2$ MK and $T \approx 1$ MK, respectively.

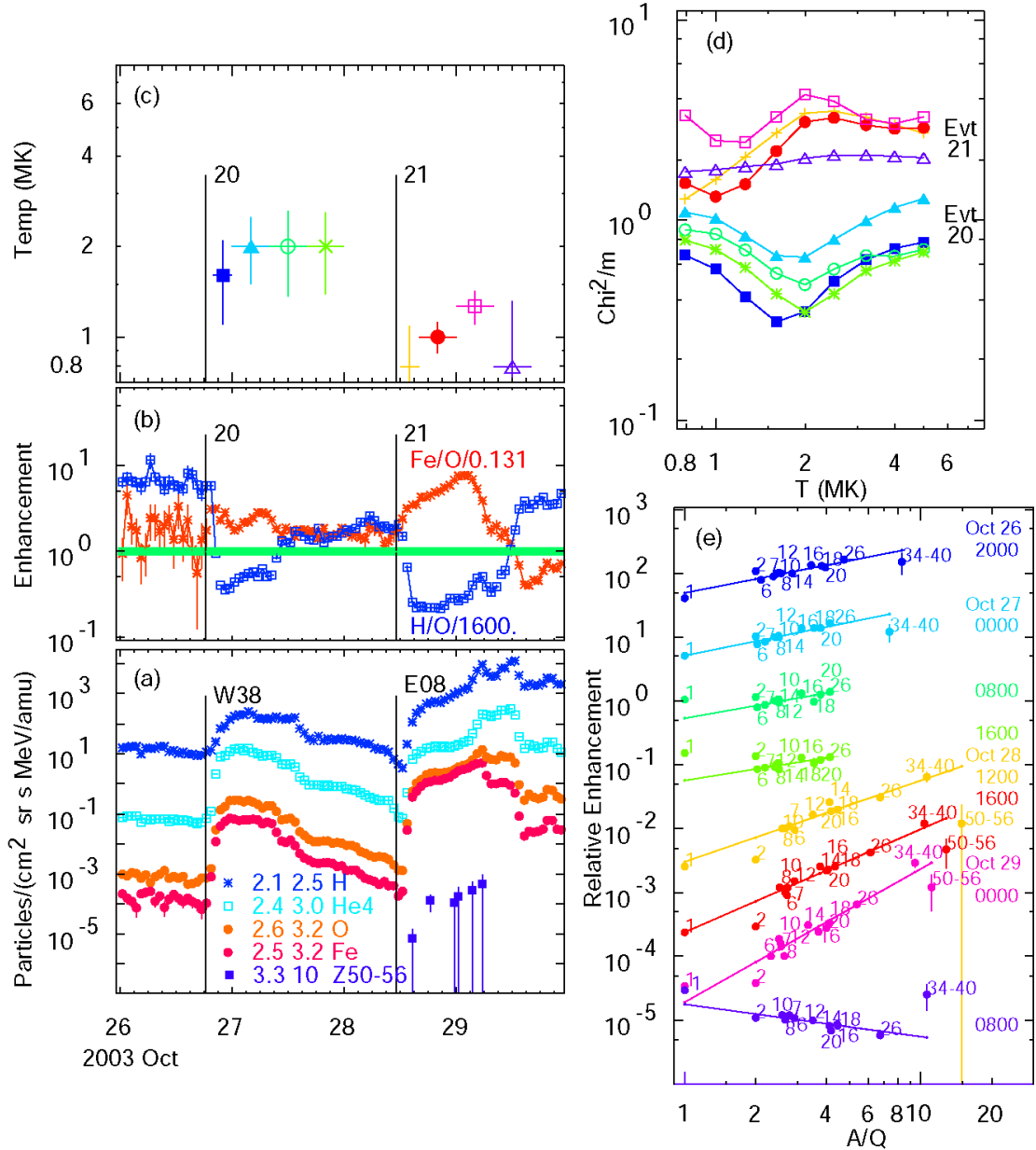

**Fig. 7** Intensities of H, He, O, Fe, and 50≤ $Z$ ≤ 56 ions (**a**), normalized abundance enhancements H/O and Fe/O (**b**), and derived source temperatures (**c**) are shown *versus* time for the 26 and 28 October 2003 SEP events. *Panel* (**d**) shows $\chi^2/m$ versus $T$ for each 8-hr interval while (**e**) shows enhancements, labeled by $Z$, *versus* $A/Q$ for each 8-hr interval shifted ×0.1, with best-fit power law for elements with $Z \geq 6$ extrapolated down to H at $A/Q = 1$. Colors correspond for the eight intervals in (**c**), (**d**), and (**e**) and symbols in (**c**) and (**d**); times are also listed in (**e**). Event onsets are flagged with solar longitude in (**a**) and event number from Reames (2016a) in (**b**), (**c**), and (**d**). H fits the power laws in $A/Q$ in these two large SEP4 events (Reames 2019c).

Low-$Z$ ions at $T \approx 3$ MK in impulsive events tend to be fully ionized, so abundance ratios like He/C or even C/O are unaltered by acceleration and tend to reflect variations in the local coronal plasma or acceleration features of each jet. These variations from one jet to another in impulsive SEPs are ≈ 30%, far in excess of statistical errors. However, when shocks reaccelerate ions from the impulsive suprathermal pools (Desai et al., 2003; Wiedenbeck et al. 2013; Bučík et al., 2014, 2015; Chen et al., 2015), fed by a large number of small impulsive events (nanojets?), the abundances are averaged over ~$n$ events so that the variations are reduced by a factor of $\sqrt{n}$. These abundance variations in $T \approx 3$ MK gradual events are ≈10%, suggesting averaging over ~10 events in the recycled impulsive pools. Statistically, we can distinguish the single-jet impulsive SEP2 events from the pool-fed gradual SEP3 events by the spread in the abundances like He/C. The abundance patterns and possible sources are shown in Fig. 8.

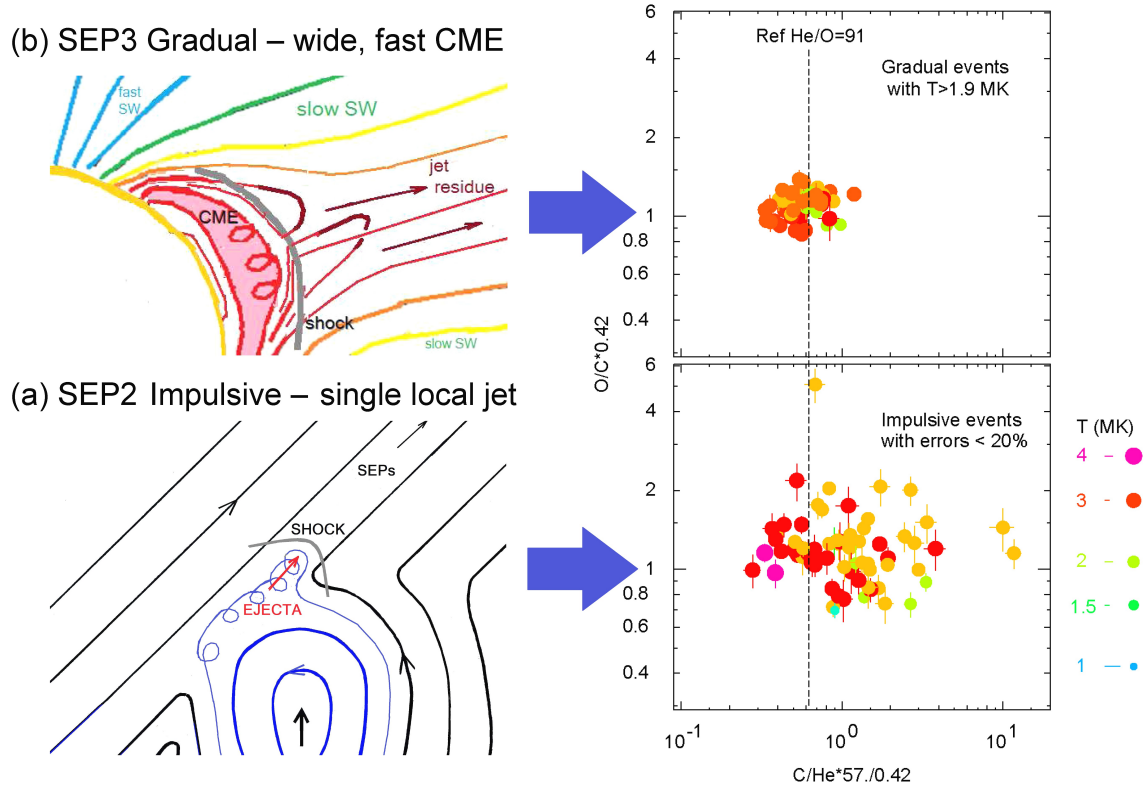

**Fig. 8 (a)** Cartoon on the *left* shows SEP2 acceleration from one of the individual jets leading to non-statistical spread in abundances in O/C vs. C/He at 2.5 to 3.2 MK for different events on the *right*. **(b)** Cartoon on the *left* shows a shock traversing a pool of multi-jet residue producing SEP3 event with $T >2$ MK with average abundances of O/C vs. C/He in each event as shown on the *right* (Reames 2016a. 2020a, 2021b). The spread in abundances is reduced by pre-event averaging in multi-jet pools of seed particles.

In Table 1 we summarize the four categories of SEP events based upon their abundance patterns as defined by Reames (2020a, 2021b). SEP1 and SEP2 are subclasses of impulsive events with and without significant shock acceleration, while SEP3 and SEP4 are both gradual SEP events dominated by different seed populations.

Table 1 Properties of the four SEP element abundance patterns.

|  | **Observed Properties** | **Physical Association** |
|---|---|---|
| **SEP1** | Power-law enhancement vs. $A/Q$ (Fe/O >×4) with $T \approx 3$ MK including $Z=1$ and $Z >2$ | Magnetic reconnection in solar jets with no fast shock |
| **SEP2** | Power-law enhancement vs. $A/Q$ (Fe/O > ×4) with $T \approx 3$ MK for $Z >2$; ~30% scatter in He/C, etc. Proton excess ~×10 | Jets with fast, narrow CMEs drive shocks that reaccelerate SEP1 ions plus excess protons from ambient plasma |
| **SEP3** | Power-law enhancement vs. $A/Q$ (Fe/O < ×4) with $T \approx 3$ MK for $Z >2$; <10% scatter in He/C, etc. Proton excess ~×10 | Moderately fast, wide CME-driven shocks accelerate SEP1 residue left by many jets in pools, plus excess protons from ambient plasma |
| **SEP4** | Power-law enhancement vs. $A/Q$ with $0.8< T < 1.8$ MK for $Z=1$ and $Z > 2$ | Extremely fast, wide CME-driven shocks accelerate all ions so that ambient plasma dominates. |

## 2.2.3 FIP

Modern FIP theory depends upon the presence of Alfvén waves that generate a ponderomotive force that helps to drive low-FIP ions across the chromosphere, but cannot affect high-FIP neutral atoms (Laming 2009, 2015). These calculations follow the detailed evolution of the ionization states, the capture and loss of electrons, of the elements as they cross the chromosphere in the presence of Alfvén waves. All elements then become highly ionized at coronal temperatures. The calculations distinguish transport along open and closed field lines, since the Alfvén waves can resonate with the loop length on closed field lines, altering the behavior (e.g. Laming et al. 2019).

Independently, detailed comparisons of abundance measurements of SEPs and of the solar wind began to show that SEPs were not simply reaccelerated solar wind (Mewaldt et al. 2002; Desai et al. 2003; Kahler and Reames 2003; Kahler et al. 2009; Reames 2018a; 2020a). In particular, the FIP value of the crossover between low and high FIP was higher in the solar wind (~14 eV) than in SEPs (~10 eV), so that the elements C, P, and S behave as low-FIP in the solar wind but high-FIP in SEPs. This suggested that the elements that will eventually become SEPs tend to cross into the corona on closed magnetic field lines, such as those in active regions, while elements that will become solar wind enter the corona on open field lines (Reames 2018a). The relative abundances and FIP patterns are compared in Fig. 9.

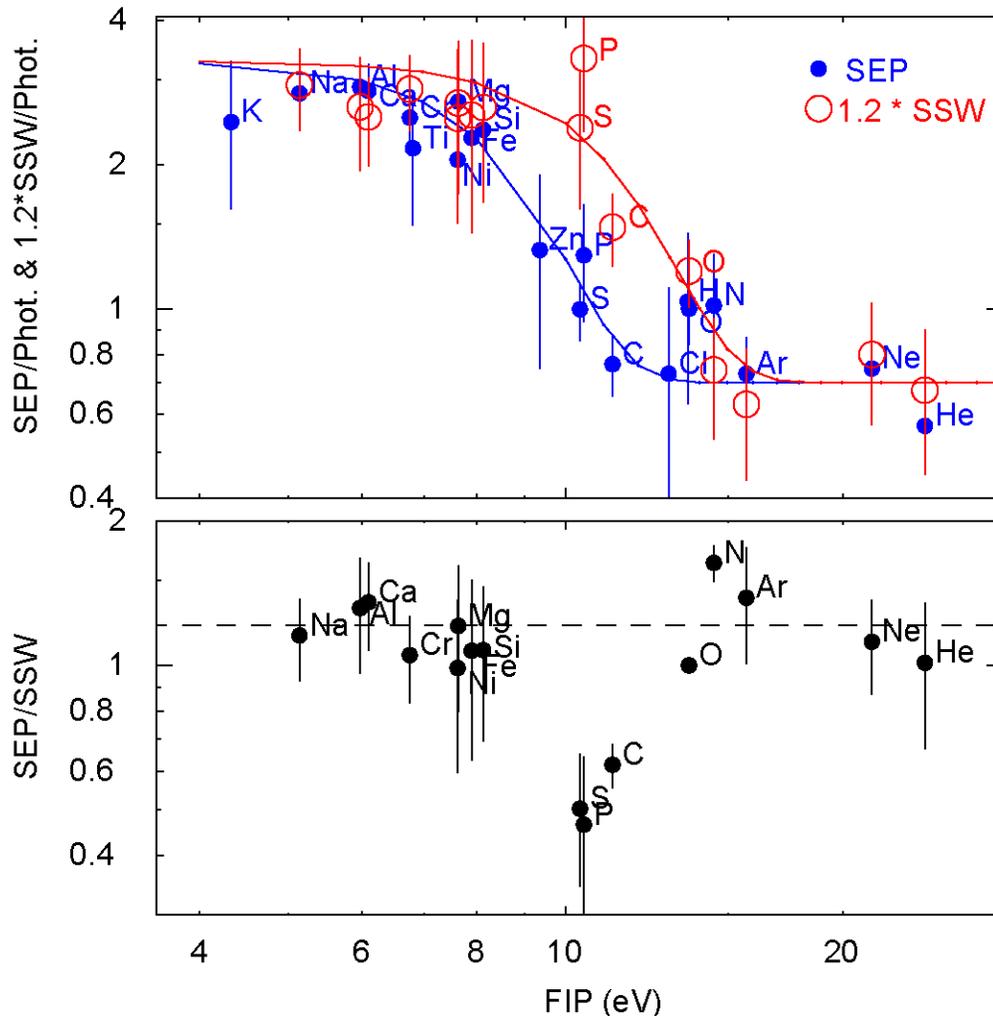

**Fig. 9** The *lower* panel shows the direct ratio of SEP to slow solar wind (SSW; Bochsler 2009) abundances, each normalized at O. The *upper* panel compares "FIP plots" of SEP and SSW abundances relative to the photosphere showing differences in the crossover from low to high FIP (Reames 2018a).

There are also differences in solar photospheric abundances used as a reference for FIP measurements. Meteorites retain a memory of abundances of non-volatile elements in the pre-solar nebula and abundances of 40 elements show <10% agreement with spectral line measurements of the photosphere (Lodders et al. 2009). Caffau et al. (2011) have supplemented these with measurements of the dominant volatile elements, resolved as necessary from blended spectral lines. In contrast, Asplund et al. (2009, 2021) have taken measurements from only well-resolved spectral lines. These abundance patterns are listed in Table 2 and SEP FIP dependences based upon them are compared in Fig. 10 along with the theoretical values expected (Laming et al. 2019).

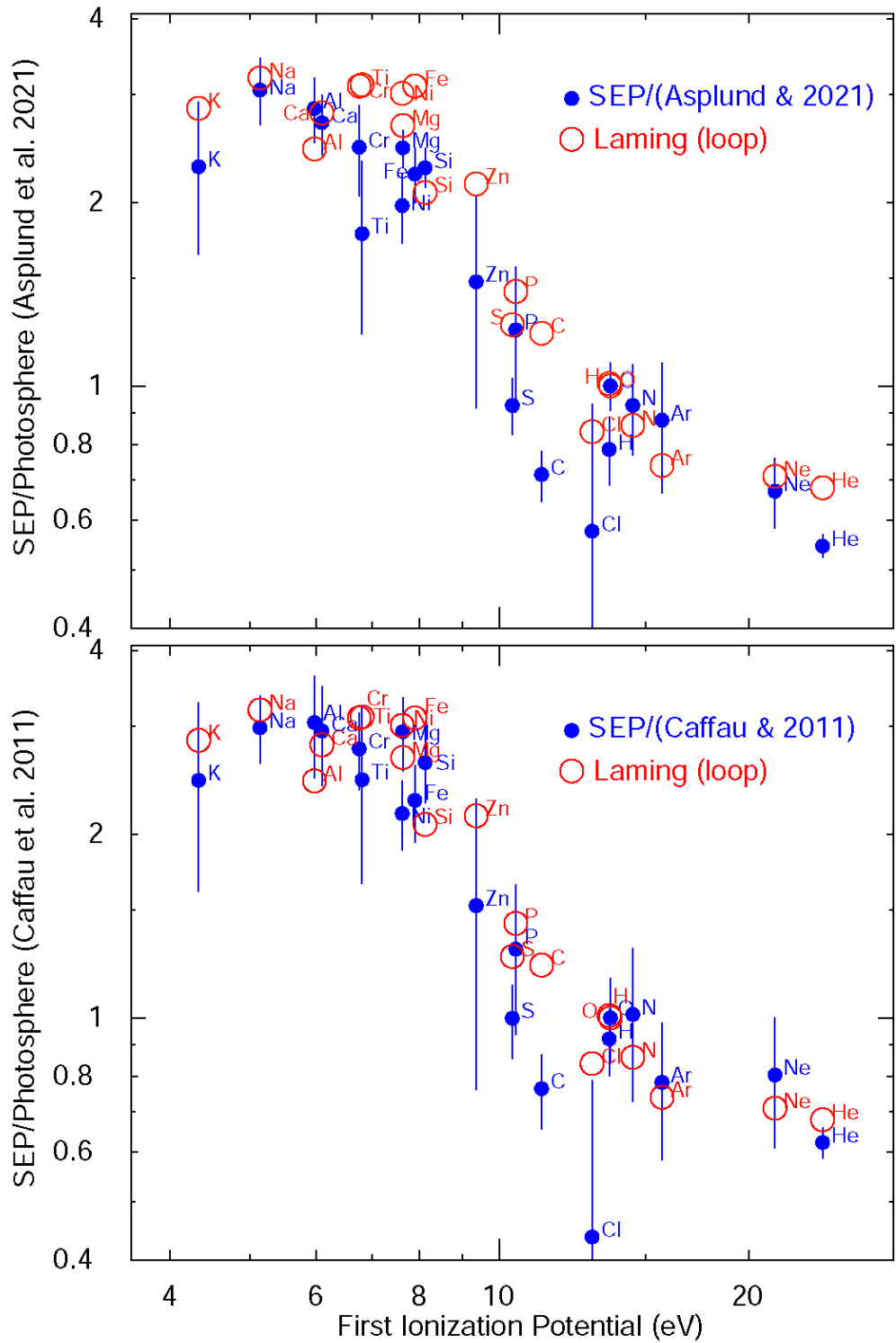

**Fig. 10** The panels compare SEP to photospheric abundance ratios vs. FIP and the theoretical prediction of Laming et al. (2019) using photospheric abundances of Caffau et al. (2011) and Lodders et al. (2009), *lower* panel, and of Asplund et al. (2021), *upper* panel.

**Table 2** Photospheric, SEP, and Slow Solar Wind Abundances.

|    | Z | FIP [eV] | Photosphere[1] (Caffau, Lodders) | Photosphere[2] (Asplund) | SEPs[3] | Interstream Solar Wind[4] |
|----|---|----------|-----------------------------------|---------------------------|---------|---------------------------|
| H  | 1  | 13.6 | $(1.74\pm0.04)\times10^{6\,*}$ | $(2.04\pm0.0)\times10^{6}$ | $(\approx1.6\pm0.2)\times10^{6}$ | – |
| He | 2  | 24.6 | $1.46\pm0.07\times10^{5}$ | $(1.67\pm0.05)\times10^{5}$ | 91000±5000 | 90000±30000 |
| C  | 6  | 11.3 | 550±76* | 589±54 | 420±10 | 680±70 |
| N  | 7  | 14.5 | 126±35* | 138±22 | 128±8 | 78±5 |
| O  | 8  | 13.6 | 1000±161* | 1000±92 | 1000±10 | 1000 |
| Ne | 10 | 21.6 | 195±45 | 234±27 | 157±10 | 140±30 |
| Na | 11 | 5.1  | 3.47±0.24 | 3.39±0.23 | 10.4±1.1 | 9.0±1.5 |
| Mg | 12 | 7.6  | 60.3±8.3 | 72.5±5.0 | 178±4 | 147±50 |
| Al | 13 | 6.0  | 5.13±0.83 | 5.50±0.38 | 15.7±1.6 | 11.9±3 |
| Si | 14 | 8.2  | 57.5±8.0 | 66.1±4.6 | 151±4 | 140±50 |
| P  | 15 | 10.5 | 0.501±0.046* | 0.525±0.036 | 0.65±0.17 | 1.4±0.4 |
| S  | 16 | 10.4 | 25.1±2.9* | 26.9±1.9 | 25±2 | 50±15 |
| Cl | 17 | 13.0 | 0.55±0.38 | 0.42±0.19 | 0.24±0.1 | – |
| Ar | 18 | 15.8 | 5.5±1.3 | 4.9±1.1 | 4.3±0.4 | 3.1±0.8 |
| K  | 19 | 4.3  | 0.224±0.046* | 0.240±0.017 | 0.55±0.15 | – |
| Ca | 20 | 6.1  | 3.72±0.60 | 4.07±0.28 | 11±1 | 8.1±1.5 |
| Ti | 22 | 6.8  | 0.138±0.019 | 0.191±0.022 | 0.34±0.1 | – |
| Cr | 24 | 6.8  | 0.759±0.017 | 0.851±0.078 | 2.1±0.3 | 2.0±0.3 |
| Fe | 26 | 7.9  | 57.6±8.0* | 58.9±5.4 | 131±6 | 122±50 |
| Ni | 28 | 7.6  | 2.95±0.27 | 3.24±0.30 | 6.4±0.6 | 6.5±2.5 |
| Zn | 30 | 9.4  | 0.072±0.025 | 0.0742±0.0085 | 0.11±0.04 | – |
| Se–Zr | 34–40 | – | ≈0.0118 | ≈0.0108 | 0.04±0.01 | – |
| Sn–Ba | 50–56 | – | ≈0.00121 | ≈0.00135 | 0.0066±0.001 | – |
| Os–Pb | 76–82 | – | ≈0.00045 | ≈0.00040 | 0.0007±0.0003 | – |

[1] Lodders et al. (2009). Ratios to O of other elements from Lodders et al. (2009) are taken before correction of O by Caffau et al. (2011).
* Caffau et al. (2011).
[2] Asplund et al. (2021).
[3] Reames (1995a, 2014, 2020a).
[4] Bochsler (2009).

### 2.2.4 Spectra

For many SEP events the element abundances are determined by one physical process, such as the abundance enhancements at magnetic-reconnection sites in jets, while the spectra are determined or greatly modified by another process, e.g. shock acceleration. Thus, they need not be related. However, for the large events we call SEP4 events, shock waves select the ions of different elements from the ambient coronal plasma and also determine their spectra. For these events we find an underlying relationship between the power-laws of abundances vs. *A/Q* and the spectral indices, although the relationship can be damaged by strong wave generation and scattering in the most intense events (Reames 2020b, 2021a). Also, for some events a single spectral index or power law in *A/Q* does not suffice for all species.

Figure 10 shows correlations among spectral indices and powers of *A/Q* for gradual SEP events. He spectra are slightly harder than O spectra in Fig. 11a, and there is a greater spread in the indices of Fe vs. O in Fig. 11b. Spectral indices of Fe and O are determined at similar values of energy per nucleon, but the magnetic rigidity of Fe is about twice that of O for these measurements, so the spectral indices vary with rigidity.

Figure 11c shows spectral indices of O vs. power-law enhancement in *A/Q*. Here, the abundances and spectral indices are decoupled for the SEP3 events (orange and red) because of the pre-enhanced seed population. However, for the remaining SEP4 events (blue and green), when abundances at constant velocity vary as $(A/Q)^x$ and spectra as $E^y$, then there is a strong correlation of $y = x/2 - 2$ (Reames 2020b, 2021a). The origin of this coupling relates to the fundamental physics of shock acceleration and to the "injection problem" (e.g. Zank et al. 2001) of sampling by the shock of its seed population, but it is not yet understood.

High-energy breaks or steepening of spectra are common in gradual SEP events (Tylka et al. 2005; Mewaldt et al. 2012; Tylka and Dietrich 2009). These break energies depend upon *A/Q*, vary between 10 and 120 MeV amu$^{-1}$, and have been studied using diffusion theory (Zhao et al. 2016; Desai et al. 2016; Li et al. 2009). Spectral breaks provide an additional way to study *A/Q* dependence.

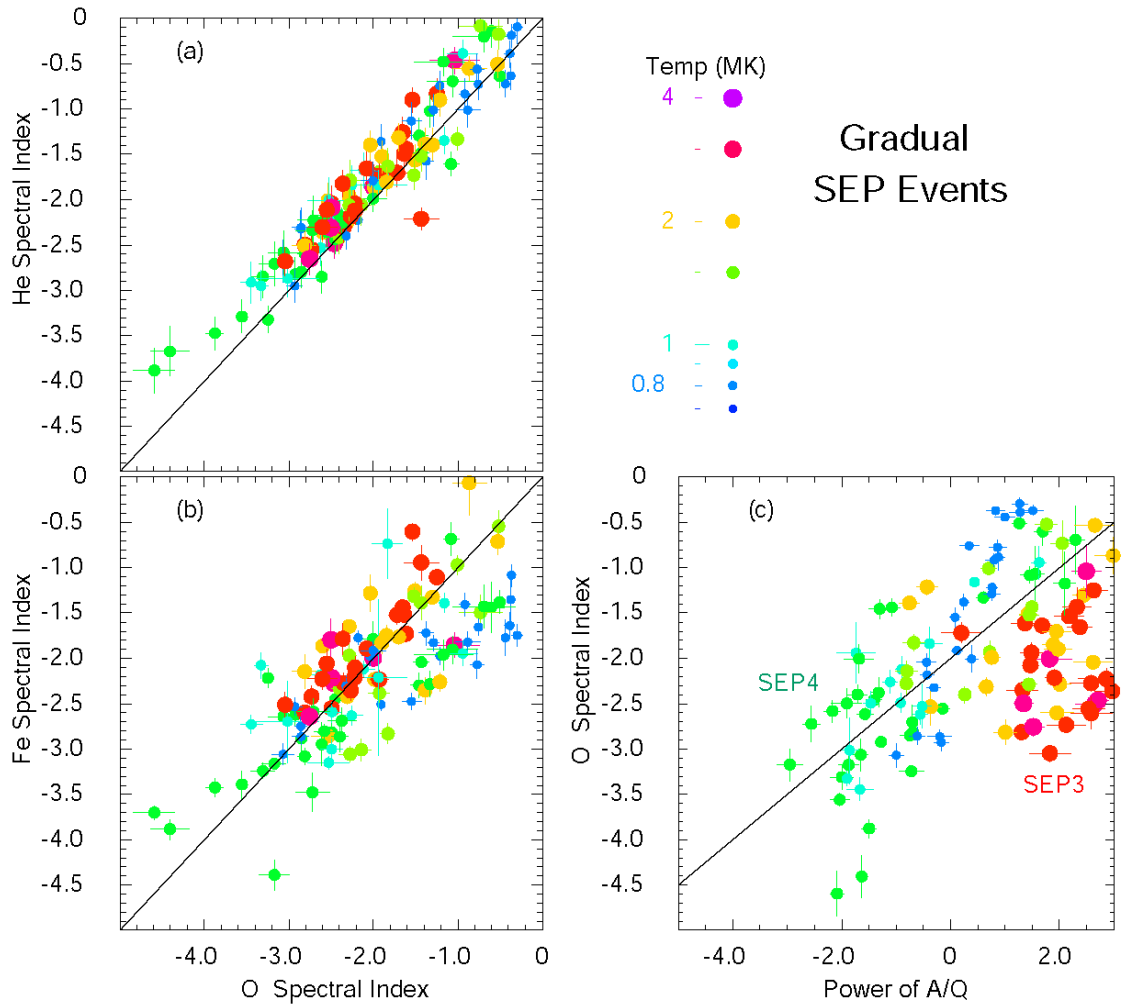

**Fig. 11** Panels show the spectral indices of **(a)** He vs. O and **(b)** Fe vs. O, and **(c)** spectral indices of O vs. the power of $A/Q$ for the first four 8-hr intervals in 45 gradual SEP events listed by Reames (2016a). Color and size of points show the power-law-derived source plasma temperature, distinguishing SEP3 events with $T \geq 2$ MK (orange, red) and pre-enhanced abundances, from the SEP4 events (green, blue). In **(a)** and **(b)**, the solid line is diagonal, $y = x$, in **(c)**, it is $y = x/2 - 2$ (Reames 2021a).

### 2.2.5 Acceleration vs. Transport

Are the element abundances seen in gradual SEP events mainly determined by the time of acceleration or strongly modified by differences in the scattering of the ions during transport to the observer? We are beginning to address this question (e.g. Reames 2020b). It is possible to rank gradual events by size: There are some small gradual SEP events that seem to propagate nearly scatter free, like impulsive SEP events. Small and moderate events tend to be acceleration dominated where abundances and energy spectra do not change much with time or space. SEP3 events are generally moderate – the abundances change little with time. However, in large SEP events, where intensities approach the "streaming limit" (Reames and Ng 1998, 2010), abundances and spectra can change markedly with time (Reames 2020b, 2021a). Nevertheless, it is difficult to call these events transport-dominated since blockage in transport also affects acceleration.

SEPs are not accelerated from the solar wind. They have a different FIP pattern that may arise from plasma near active regions, sampled at ~1.5 $R_S$ in impulsive events and ~2 – 3 $R_S$ in gradual events, which may be maintained and reaccelerated by shocks as they move away from the Sun.  As the shocks move radially and the Sun rotates, samples from neighboring longitudes in the Parker spiral, neighboring flux tubes, are coupled together by scattering at each point on the shock.  By the time it reaches 1 AU, each point on the shock has crossed and sampled field lines spanning ≈80⁰ of solar longitude.  "Event averages" represent large spans and no longer say anything about possible variations along the shock.  Perpendicular diffusion at locations away from the shock is usually irrelevant.

## 3  Answers Leading to Questions

It took us nearly 30 years to establish that gradual SEP events did not come from flares; the biggest SEP events were accelerated by CME-driven shock waves.  In recent years we have found that even the impulsive SEP events do not come from flares; they come from solar jets – a second paradigm shift.  SEP source temperatures never reach ~10 MK; at most they reach 2 – 3 MK.

But now the study of abundances of elements in SEP events has grown into a rich new field that has allowed us to probe many different physical processes that occur in the solar corona.  Some of these processes would be invisible otherwise.  Yet this field has left lingering questions from the early days and has also raised some new ones.

### *3.1 Theory of Impulsive SEPs*

$^3$He enhancements and heavy-ion enhancements each have their own explanation, but how do they fit together?  Collapsing islands of magnetic reconnection could easily provide the mirroring ions that absorb waves, as envisioned by Temerin and Roth (1992) but do streaming electrons generate those waves?  Can we explain the rare, small events with huge, apparently resonant, increases of Si, or S below 1 MeV amu$^{-1}$ (Mason et al. 2016; Reames 2021b)? How do the spectral determinants of $^3$He (e.g. Liu et al. 2006) apply to Si, S, or Fe?  Can essentially all these elements in the active volume be accelerated by resonant absorption of waves at $A/Q = 3$, as is the case for $^3$He with $A/Q = 1.5$, while H and $^4$He damp the waves that resonate at $A/Q = 1$ or 2?

### *3.2 Spectra vs. Abundances*

It seems reasonable that there should be a correlation between energy spectral indices and powers of $A/Q$ in shocks, but the precise numerical relationship of the exponents, $y = x/2 – 2$, was not predicted in advance.  Models of SEP acceleration and transport generally do not make use of the well-defined coronal abundances input to predict the abundance variations output, so, without modeling, the variations are difficult to study.  To what extent do the abundance patterns depend upon shock properties?

We would not expect correlations between spectra and abundances for SEP2 or SEP3 events since the spectra are determined by shocks and the abundances are determined by magnetic reconnection or wave resonance.  However, why is there no correlation between spectra and abundances in impulsive SEP1 events (Reames 2021a)?

How do different aspects of magnetic reconnection determine the spectra and the abundances independently?

### *3.3 Variations in FIP – He – C*

The photospheric abundances do affect the agreement of SEPs with theoretically expected (Laming et al. 2019) values. Caffau et al. (2011) values seem to be in better agreement both at He and at low FIP, than Asplund et al. (2021) values.

Breneman and Stone (1985) concluded that their average SEP abundances had a net negative slope in $A/Q$ that needed to be corrected upward to obtain coronal abundances. Later studies with averages of larger numbers of events found corrections unnecessary. In effect, they probably used SEP3 events with positive powers of $A/Q$ to compensate for SEP4 events with negative powers. We cannot exclude a small correction that would raise Fe/O by 5 – 10%, but it is not clearly required. However, we cannot use such a correction to raise the abundances of both Fe and He while keeping O, Mg, and Si constant. Hence, a power-law correction cannot improve agreement in the upper panel of Fig. 10.

The abundance of He shows event-to-event variations in both impulsive and gradual events which is often suggested to be related to the slow ionization of He because of its very high FIP=24.6 eV, but Ne at 21.6 eV does not share these variations at all. Impulsive events with 3.2 MK source temperatures, where $A/Q$ variations are precluded, have source He/O ≈ 90, as do SEP3 events that reaccelerate these ions. Despite this, a few rare small SEP1 events have strongly suppressed values as low as He/O ≈ 2 (Reames 2019a); we do not understand these events. Are there sometimes He-poor jets? Most SEP4 gradual events have He/O ≈ 50, not 90 (Reames 2017). Why? Where do they sample the ambient coronal plasma? It seems unlikely that these events have $A/Q$-enhanced O.

The ratio of C/O = 0.420 ± 0.010 in SEPs is uniquely below theoretical expectations as is the ratio in the solar wind (e.g. Reames 2020a). Could the photospheric C/O = 0.589 ± 0.054 be 40% too high? The SEP C/O is always below 0.5, in every event, well below the photospheric value. Is there some mechanism that selectively suppresses C in the corona, or in the SEPs? Older, e.g. Anders and Grevesse (1989), photospheric values of C/O = 0.489 are in better agreement with SEPs.

60 Years									D. V. Reames